\documentclass[preprint]{revtex4-1} 

\usepackage{graphicx} 
\usepackage{hyperref}

\begin{document}

\title{Design of simple stellarator using tilted toroidal field coils}

\author{Yasuhiro Suzuki}
\email{suzuki.yasuhiro@nifs.ac.jp}
\affiliation{National Institute for Fusion Science, National Institutes of Natural Sciences, Toki 5095292, Japan}
\affiliation{The Graduate University for Advanced Studies, SOKENDAI, Toki 5095292, Japan}
\author{Jie Huang}
\affiliation{National Institute for Fusion Science, National Institutes of Natural Sciences, Toki 5095292, Japan}
\author{Nengchao Wang}
\affiliation{International Joint Research Laboratory of Magnetic Confinement Fusion and Plasma Physics, State Key Laboratory of Advanced Electromagnetic Engineering and Technology, School of Electrical and Electronic Engineering, Huazhong University of Science and Technology, Wuhan, 430074, China}
\author{Yonghua Ding}
\affiliation{International Joint Research Laboratory of Magnetic Confinement Fusion and Plasma Physics, State Key Laboratory of Advanced Electromagnetic Engineering and Technology, School of Electrical and Electronic Engineering, Huazhong University of Science and Technology, Wuhan, 430074, China}

\begin{abstract}
This paper deals with the design of the stellarator field with the simple coil set. In order to realize the stellarator field by the simple coil set, the tilted toroidal field coil uses for creating the rotational transform. Sixteen tilted TF coils create the small radial field and large vertical field. With reducing the vertical field of the tilted toroidal field coil by the axisymmetric poloidal field coil, the stellarator field can be made. The formation of clear and nested flux surfaces is confirmed, and the rotational transform is proportional to the tilting angle of the toroidal field coil. Main components of the magnetic field for the simple stellarator in this paper are the large mirror ripple and small helical ripple. This is a similar property to the quasi-isodynamic configuration like the Wendelstein 7-X stellarator. The collisionless orbit for the proton is studied. For a moderate tilting angle of the toroidal field coil, the confinement of the passing and trapped particles improves. 
\end{abstract}

\maketitle

\section{Introduction}

The tokamak is the first candidate of the fusion reactor. However, although the tokamak has the good confinement, the current driven instability disrupting the plasma is a critical problem~\cite{JohnWesson2011}. Also, for the high confinement mode like the H-mode, the MHD instability in the high confinement region appears, for example, the edge localized mode (ELM) on the pedestal~\cite{Zohm1996}. The high-energy flux driven by the ELM damages the material of the divertor, and difficulties appear for the realization of the steady state fusion reactor~\cite{Loarte2003}. On the other hand, the stellarator~\cite{Spitzer1958} is an alternative. The stellarator has many advantages for the steady state operation, stable detachment, and so on~\cite{Boozer2008,Boozer2020}, but the transport property, which is comparing with the tokamak, is not good due to the complex 3D magnetic field structure, in particular, for the conventional stellarator~\cite{Ho1987}. For improving those problems, optimized stellarators are designed, and the confirmation of the optimization are examined experimentally and theoretically~\cite{Grieger1989,Anderson1995,Klinger2017}. In many cases, the optimized stellarator uses the poloidal modular coil with very sophisticated coil shape~\cite{0029-5515-22-7-001,Merkel1987}. Therefore, a simple question that those sophisticated coil shape can be extended and used in the future reactor is very critical. In the history of stellarator researches, the design using the simple coil set, the so-called simple stellarator, had considered~\cite{Funato1983,Bykov1989,Todd1990,Pedersen2004,Clark2014}. The simple stellarator is classified as a conventional stellarator, and the magnetic field is not optimized. However, the construction is easy comparing with the fully optimized stellarator, and it will be a good candidate to start up the stellarator research.

This study deals with the design of a simple stellarator using tilted toroidal field (TF) coils. The tilted TF coil produces not only the toroidal field but also small radial field and large vertical fields. This small radial field is a reason to make an error field, but that field can be used to make the rotational transform. The several stellarator using the tilted TF coils were proposed and constructed as the laboratory device. However, in previous study, the generation of clear and nested flux surfaces were confirmed,but the properties of the magnetic field was not studied in details  yet. Also, as pointed out in above, the simple stellarator has a great advantage to start up the stellarator research. In particular, the facility of the tokamak device can be used to the stellarator, and there is a further possibility to realize the hybrid operation of the stellarator and tokamak. This leads the synergy by the 3D physics. In the next section, we discuss a design of the simple stellarator by the tilted TF coils. The device size and the number of TF coils refer the J-TEXT tokamak~\cite{Ge2009}. We then study the formation of the flux surfaces, the rotational transform, and the spectrum of the magnetic field on Boozer coordinate~\cite{Boozer1982}. Also, the confinement of ions for the bulk plasma energy is studied to understand the transport property briefly. In the last section, we summarize this study.

\section{Realization of stellarator field by tilted toroidal field coils}

\subsection{Vacuum magnetic field by a tilted toroidal field coil}

\begin{figure}[htbp]
 \begin{center}
  \begin{tabular}{cc}
   (a) square shaped TF coil &
   (b) tilting angle, $\theta$, \\
   \includegraphics[height=5cm]{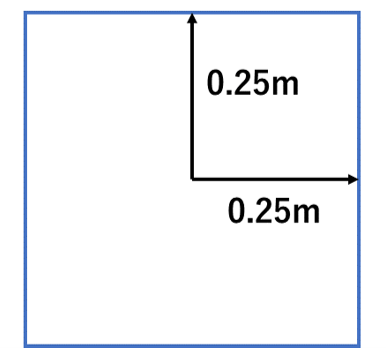} &
   \includegraphics[height=5cm]{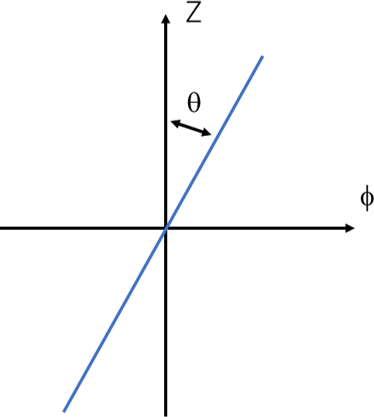}
  \end{tabular}
 \end{center}
 \caption{A schematic view of a tilted toroidal field (TF) coil for (a) $R$-$Z$ plane and (b) $\phi$-$Z$ space.}
 \label{fig:fig1}
\end{figure}

\begin{figure}[htbp]
 \begin{center}
  \includegraphics[height=5cm]{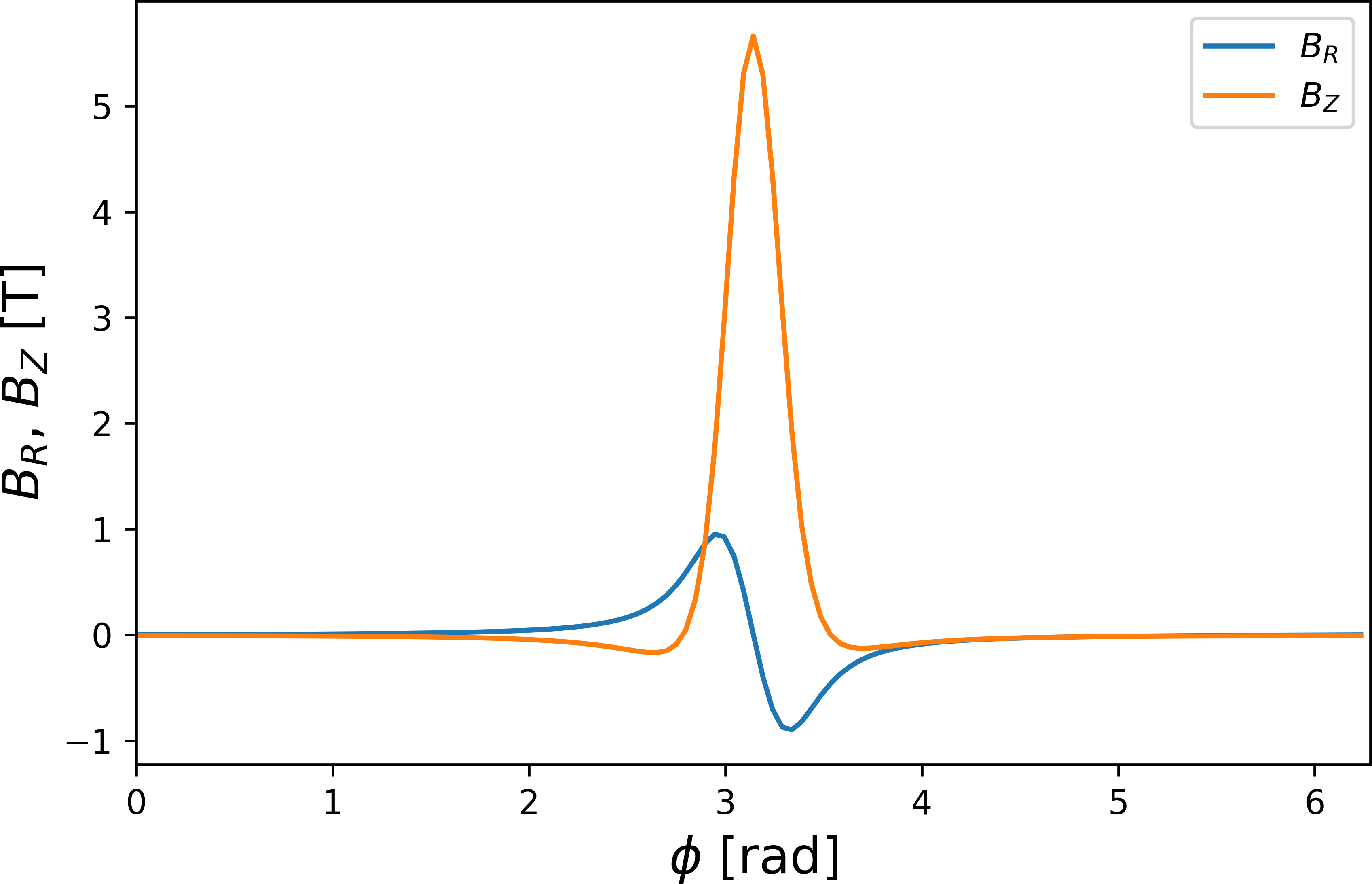}
 \end{center}
 \caption{Profiles of the radial field ($B_R$) and vertical field ($B_Z$) produced by one tilted TF coil. The toroidal field, $B_0$, is 1 T at the center of the TF coil, $R_0$ = 1 m.}
 \label{fig:fig2}
\end{figure}

Here, the vacuum magnetic field produced by a tilted TF coil is considered. Figure~\ref{fig:fig1} shows a configuration of a tilted TF coil. The size of the TF coil is 0.5 m by 0.5 m and the shape is the square on the $R$-$Z$ plane. The center of the squared TF coils is $R_0$ = 1 m. The size of the TF coil is decided from the vessel design of the J-TEXT tokamak. The tilting angle, $\theta$, is 30 degree along the toroidal angle, $\phi$. Figure~\ref{fig:fig2} shows profiles of $B_R$ and $B_Z$ along the toroidal angle, $\phi$ at $R_0$ = 1 m. The toroidal magnetic field is $B_0$ = 1 T at $R_0$ = 1 m, in other words, the current of the TF coil is estimated from the condition of $R_0 B_0$ = 1 mT. The tilting of the TF coil produces the both of $B_R$ and $B_Z$. In particular, phases of $B_R$ and $B_Z$ are opposite, and this is a reason why the tilting of the TF coil can be a source of the error field to make the magnetic island in tokamak plasmas. The amplitude of $B_Z$ is much higher than the amplitude of $B_R$. If the amplitude of $B_Z$ can be reduced by the axisymmetric poloidal field (PF) coil up to the same order of $B_R$, the vacuum magnetic field by th tilted TF coil and PF coils is equivalent to the stellarator filed produced by the helical coil.       

\subsection{A simple stellarator with tilted toroidal field and poloidal field coils}

\begin{figure}[htbp]
 \begin{center}
  \includegraphics[height=8cm]{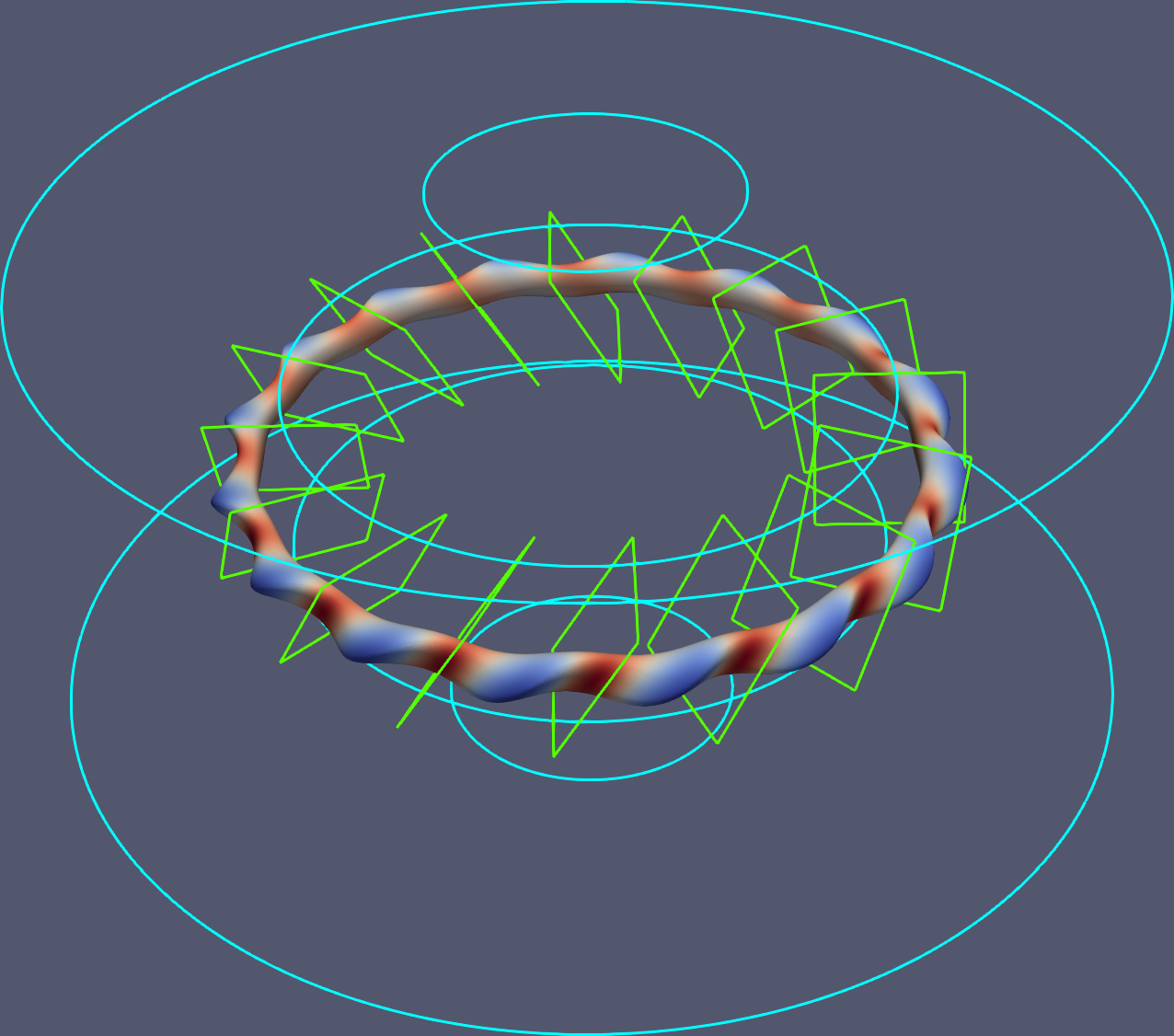}
 \end{center}
 \caption{A schematic view of a simple stellarator configuration. This configuration consists of sixteen TF coils (light green rectangles) and three pairs of axisymmetric PF coils (light blue curves). For a reference, the last closed flux surface for $\theta$ = 20 degree is also shown. The major radius of the TF coil is 1.0 m, and the toroidal field, $B_0$, is 1 T at $R_0$ = 1 m.}
 \label{fig:fig3}
\end{figure}

As discussed in above, the tilted TF coil can make the stellarator field without the toroidal current. Here, some reference designs are discussed. Figure~\ref{fig:fig3} shows a schematic view of a coil geometry. This configuration consists of sixteen tilted TF coils (light green rectangles) and three pairs of axisymmetric poloidal coils (light blue curves). For a reference, the shape of the last closed flux surface of the plasma is also shown in the figure (colors indicate the magnetic field strength). The major radius of the TF coil is 1.0 m, and the size of the square is 0.5 m by 0.5 m. The $R_0 B_0$ is 1.0 mT, that is, the toroidal magnetic field, $B_0$, at the center of the helical coil, $R_0$ = 1 m, is 1 T along the counterclockwise direction. The size and location of PF coils are also refereed from the J-TEXT tokamak.

\begin{figure}[htbp]
 \begin{center}
  \begin{tabular}{c}
   (a) $\theta$ = 10 degree \\
   \includegraphics[height=4cm]{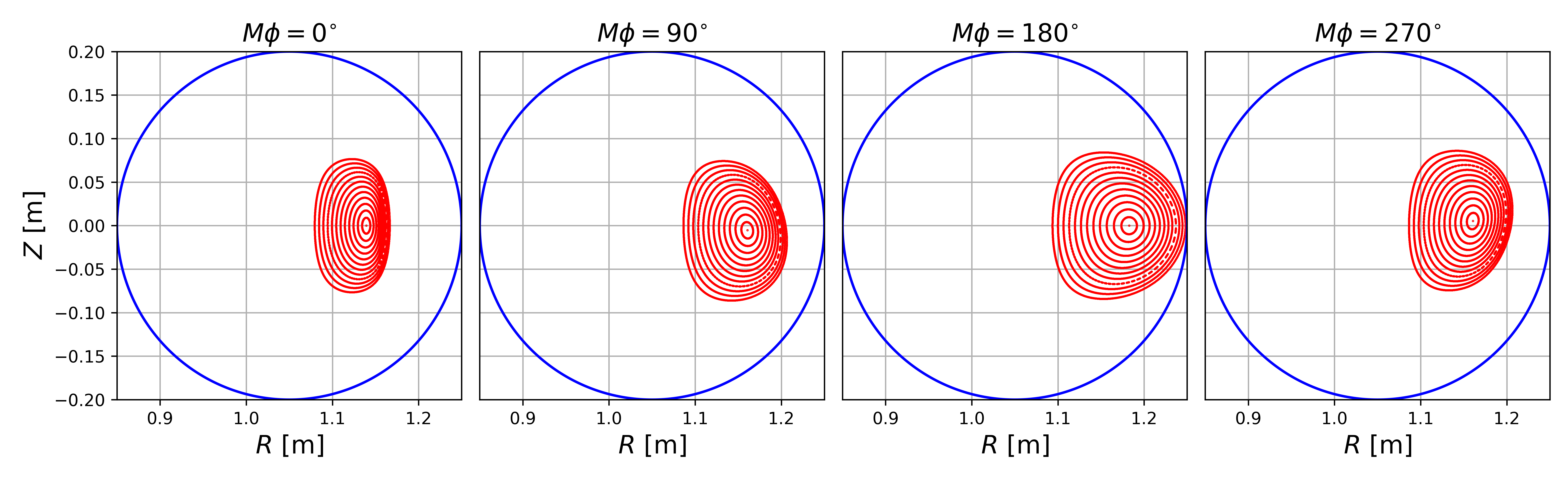} \\
   (b) $\theta$ = 20 degree \\
   \includegraphics[height=4cm]{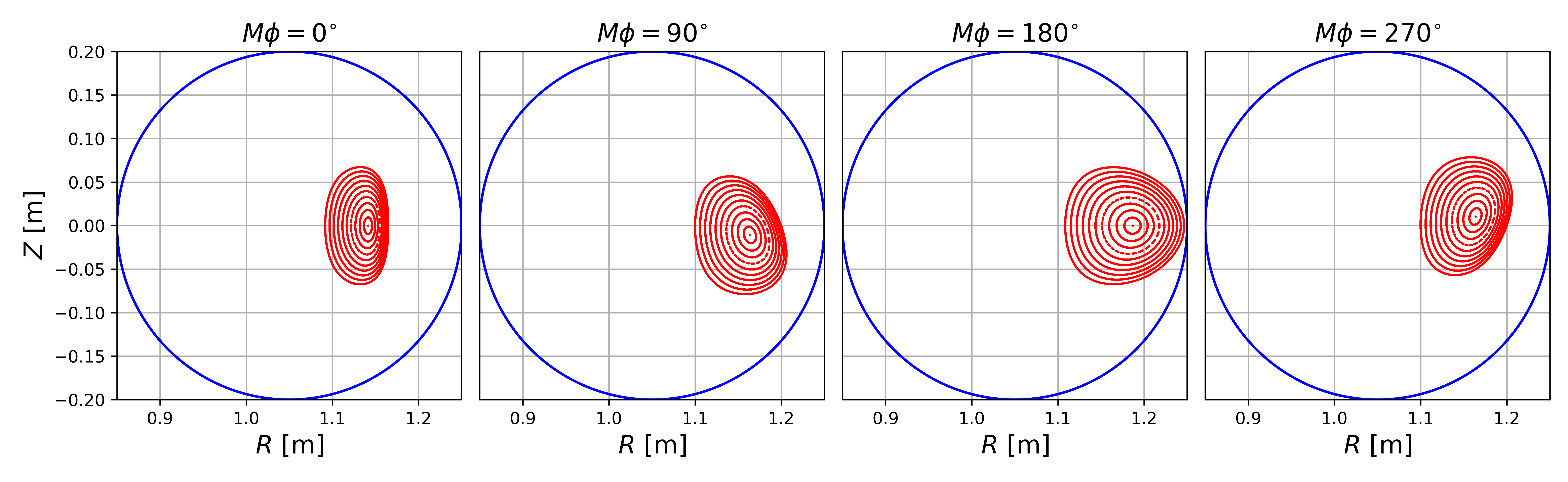} \\
   (c) $\theta$ = 30 degree \\
   \includegraphics[height=4cm]{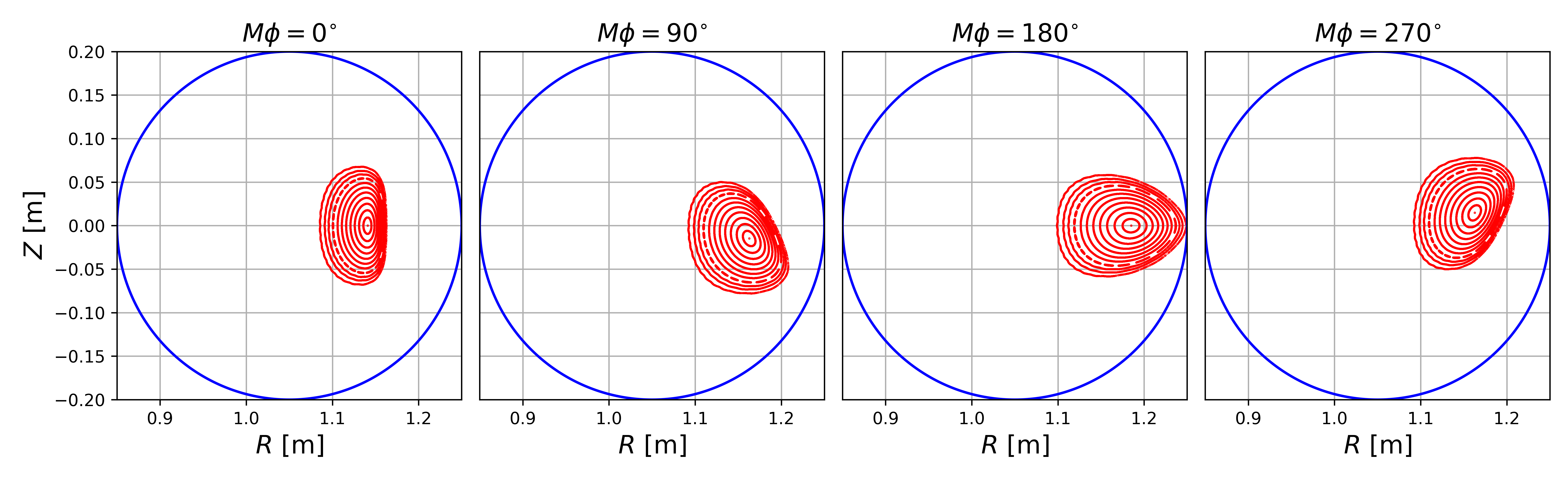}
  \end{tabular}
 \end{center}
 \caption{Poincar\'e plots of simple stellarators for different tilting angles, $\theta$ = (a) 10, (b) 20, and (c) 30 degree.}
 \label{fig:fig4}
\end{figure}

\begin{figure}[htbp]
 \begin{center}
  \includegraphics[height=5cm]{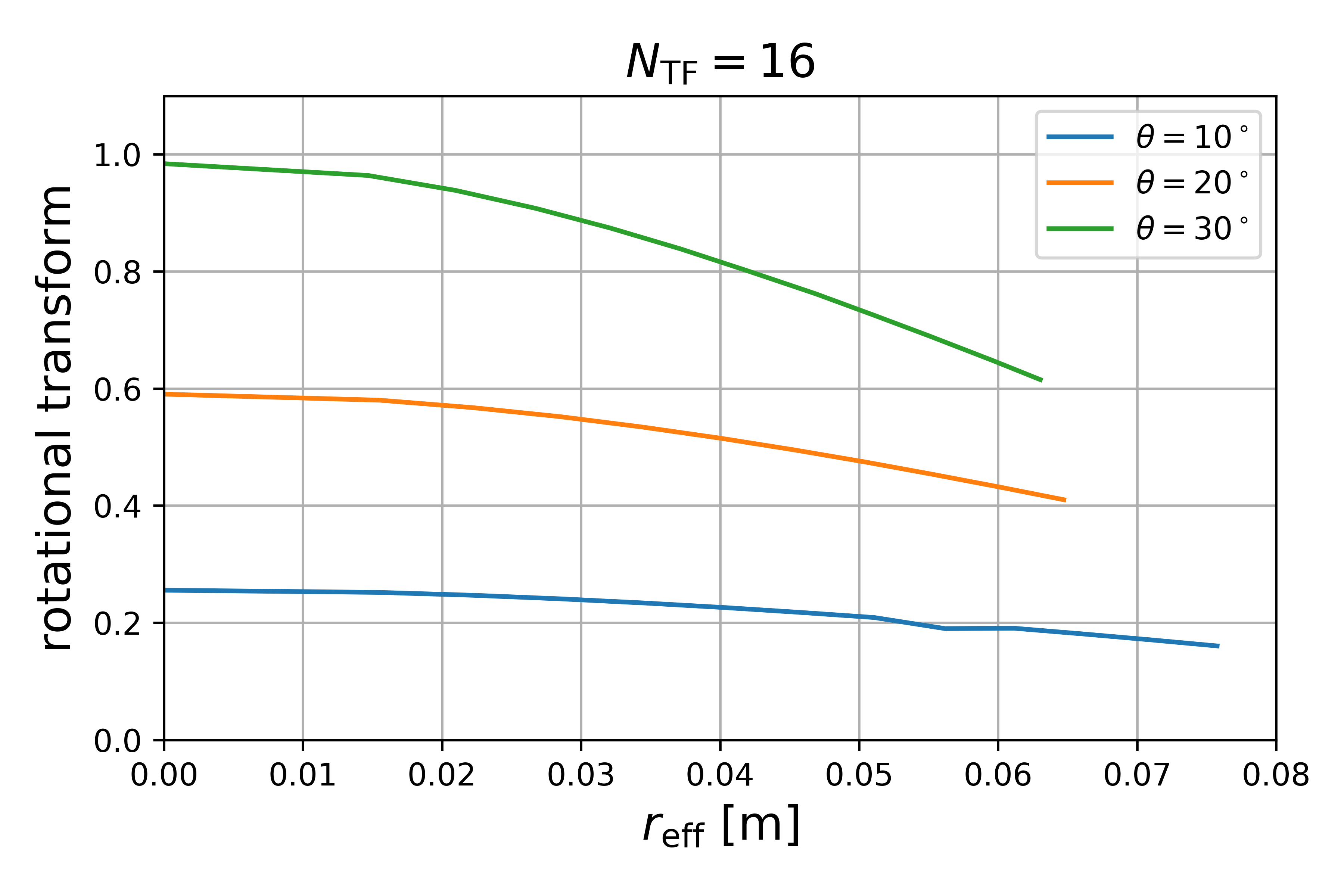}
 \end{center}
 \caption{Profiles of the rotational transform, $\iota$, for different tilting angles, $\theta$ = (a) 10, (b) 20, and (c) 30 degree.}
 \label{fig:fig5}
\end{figure}

Figure~\ref{fig:fig4} shows Poincar\'e plot of the vacuum magnetic field with different tilting angle, $\theta$, to 10, 20, and 30 degree. A blue circle in each figures indicates the first wall. As shown in figures, clear flux surfaces appear, and the first wall limits the plasma volume, that is, this configuration is the limiter configuration. For each cases, the magnetic axis is helically deviates, and the deviation is proportional to the tilting angle, $\theta$. In figure~\ref{fig:fig5}, profiles of the rotational transform, $\iota$, as the function of the effective minor radius, $r_\mathrm{eff}$, are plotted as configurations corresponding to figure~\ref{fig:fig4}. For a case of $\theta$ = 10 degree, the rotational transform is very low ($\sim$ 0.2), and the magnetic shear is very weak. With increased the tilting angle, $\theta$, the rotational transform increases, because the $B_R$ and $B_Z$ increase. For another case of $\theta$ = 30 degree, the rotational transform on the magnetic axis increases up to the unity, and gradually decreases. In that case, the magnetic shear becomes strong. However, as shown in the figure, the plasma volume shrinks, because of the large helical deviation. 
 
\subsection{Magnetic field spectrum on Boozer coordinate}

\begin{figure}[htbp]
 \begin{center}
  \begin{tabular}{ccc}
   (a) $\theta$ = 10 degree &
   (b) $\theta$ = 20 degree &
   (c) $\theta$ = 30 degree \\
   \includegraphics[height=3cm]{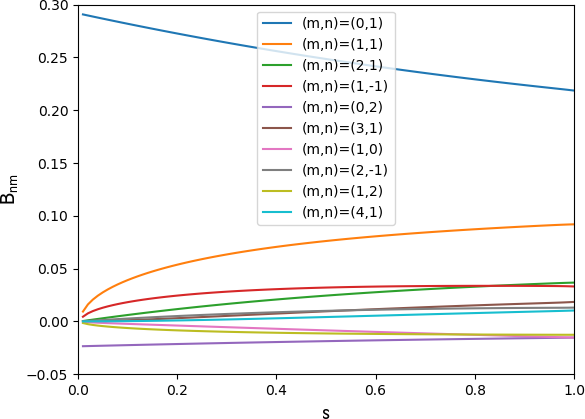} &
   \includegraphics[height=3cm]{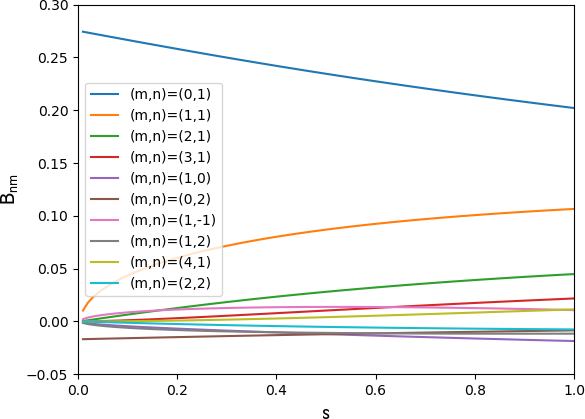} &
   \includegraphics[height=3cm]{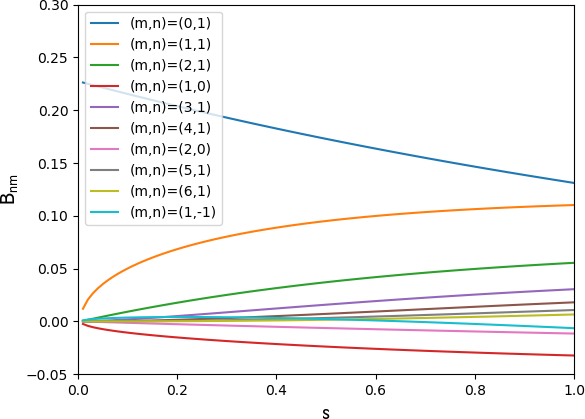}
  \end{tabular}
 \end{center}
 \caption{Spectra of the magnetic field on Boozer coordinate for different tilting angles, $\theta$ = (a) 10, (b) 20, and (c) 30 degree.}
 \label{fig:fig6}
\end{figure}

In order to understand characteristics of the magnetic configuration, the spectrum of the magnetic field is studied on Boozer coordinate~\cite{Boozer1982}. The magnetic field spectrum on Boozer coordinate is calculated from the zero-beta 3D equilibrium by the VMEC code~\cite{Hirshman1983}. Figure~\ref{fig:fig6} shows the spectrum of the magnetic field, $|B|$, on Boozer coordinate for $\theta$ = 10, 20, and 30 degree, respectively. The spectra omit the axisymmetric toroidal field component, $(n,m) = (0,0)$. Since the magnetic field is mainly produced by the TF coil, the first dominant component is the mirror ripple (TF ripple), $(n,m) = (1,0)$. The second dominant component is the helical ripple, $(n,m) = (1,1)$. This is a similar nature to the quasi-isodaynamic configuration like W7-X. With increased the tilting angle, $\theta$, the mirror ripple decreases, and the helical ripple increases. For a case of $\theta$ = 30 degree, the mirror and helical ripples are almost comparable in the plasma edge. That might reflect that the orbit property of the ion confinement will degrade because of the helical ripple loss.  Other components are relatively small comparing with the mirror and helical ripples. These spectra are studied for only the vacuum field. For cases of $\theta$ = 10 and 20 degree, since the rotational transform is small, the Shafranov shift due to the finite-$\beta$ effect might be large. An impact of the finite-$\beta$ effect on Boozer spectrum must be studied. 
 
\subsection{collisionless orbit property}

\begin{figure}[htbp]
 \begin{center}
  \begin{tabular}{ccc}
   (a) $\theta$ = 10 degree &
   (b) $\theta$ = 20 degree &
   (c) $\theta$ = 30 degree \\
   \includegraphics[height=3cm]{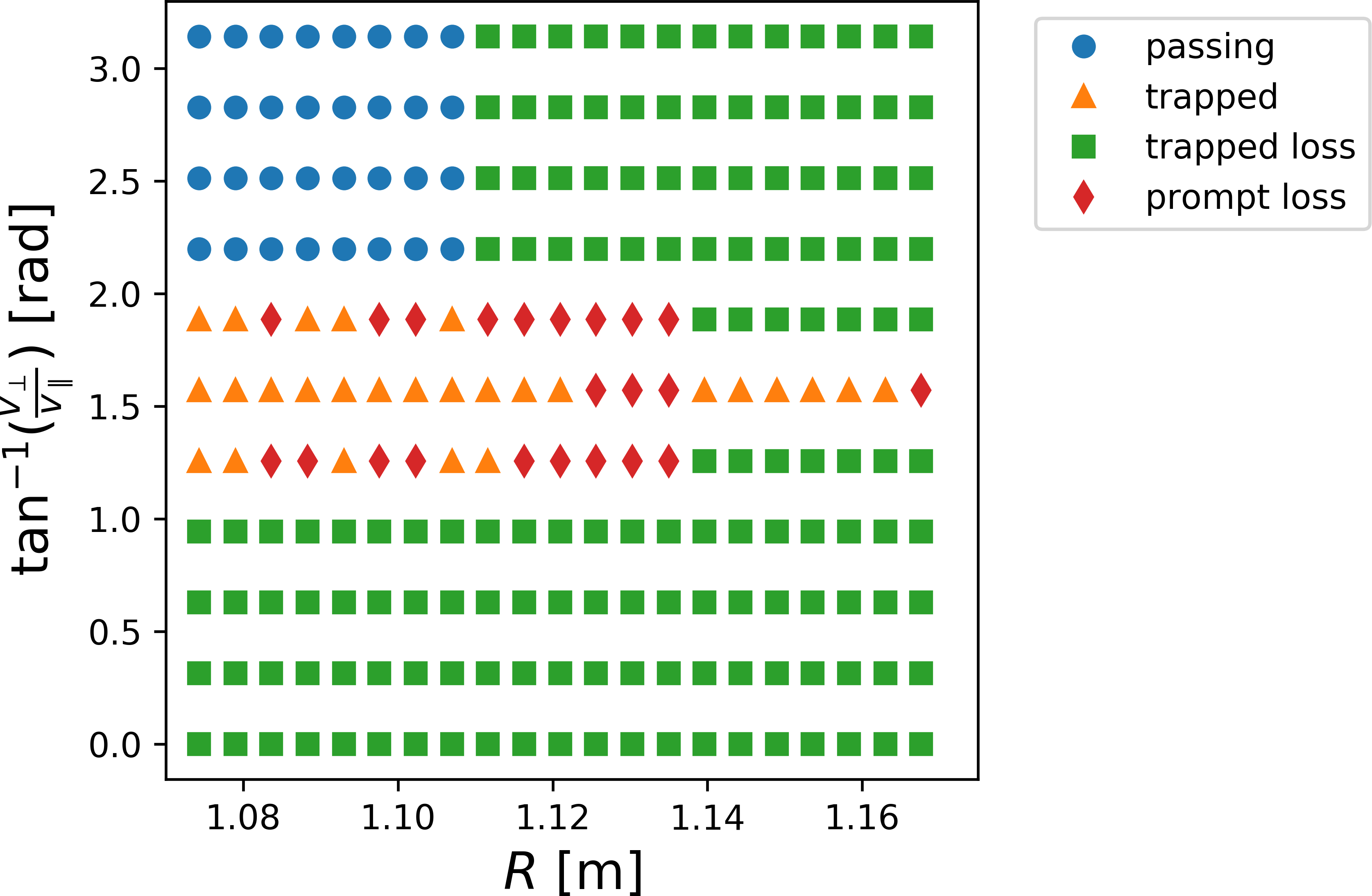} &
   \includegraphics[height=3cm]{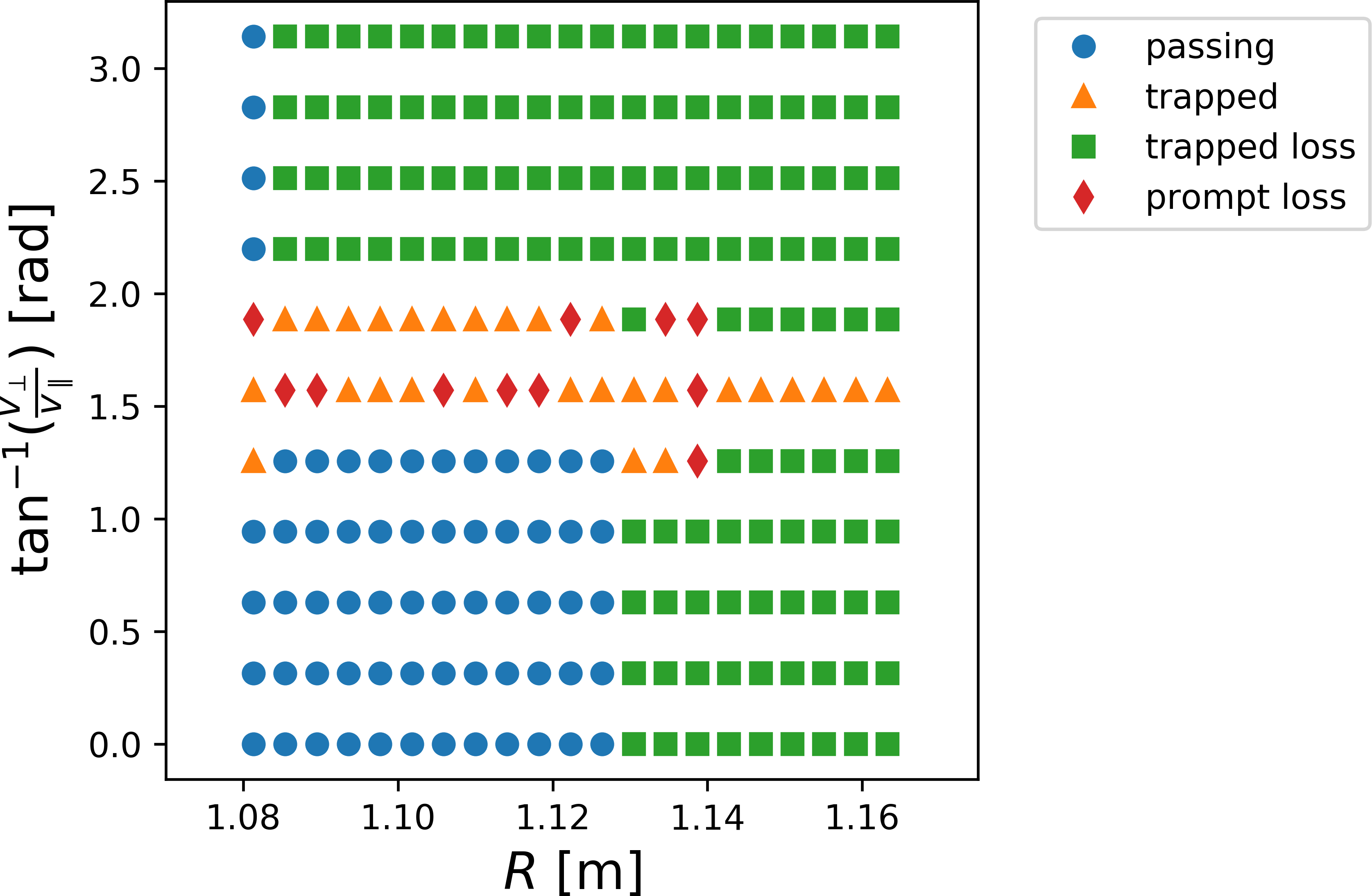} &
   \includegraphics[height=3cm]{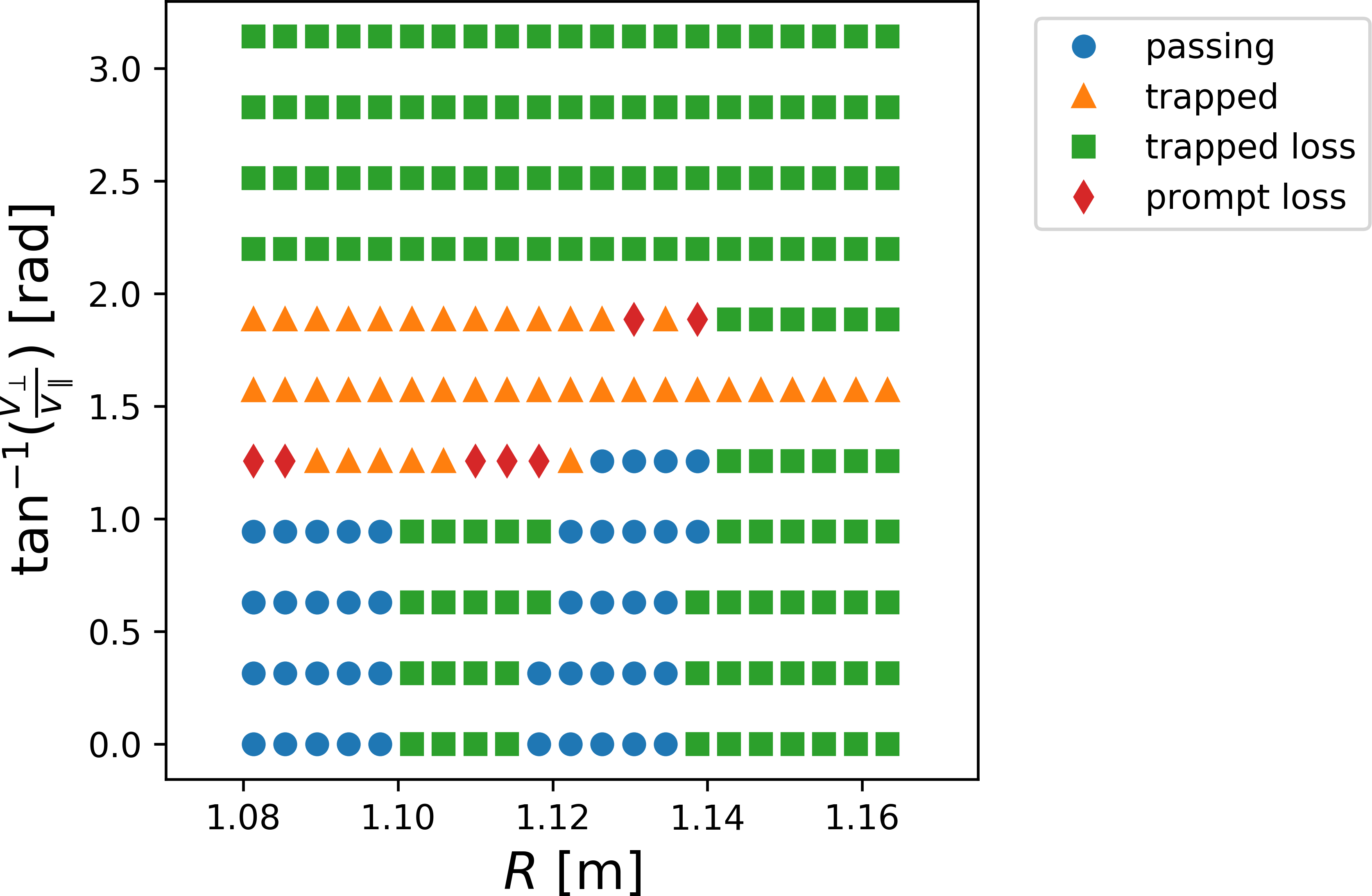}
  \end{tabular}
 \end{center}
 \caption{Loss cone of 100 eV proton orbits for different tilting angles, $\theta$ = (a) 10, (b) 20, and (c) 30 degree. The orbit following starts from Z = 0 plane with the different pitch angle.}
 \label{fig:fig7}
\end{figure}

According to the spectrum of the magnetic field on Boozer coordinate, these configurations have a similar nature to the quasi-isodynamic configuration, that is, the confinement of the trapped particle improves due to the strong mirror ripple. On the other hand, for the small tilting angle, $\theta$ = 10 degree, the rotational transform is very small. In such a case, the poloidal drift width is very large. Therefore, the orbit surfaces deviates from the flux surface, and then, the prompt loss will increase. To confirm that hypothesis and estimate the orbit confinement quantitatively, the collisionless guiding-center orbit is studied for three cases. The guiding-center of the 100 eV proton is followed by the GCR code~\cite{Suzuki2006} for the vacuum field. The starting points are set to the $Z$ = 0 plane at the $\phi$ = 0. The pitch angle, $\tan^{-1} (v_{\perp}/v_{\parallel})$, is scanned from 0 to $\pi$. Figure~\ref{fig:fig7} shows the loss cone for three cased of $\theta$ = 10, 20, and 30 degree, respectively. For a case of $\theta$ = 10 degree, due to the large poloidal drift width, the almost co-going orbit losses. The trapped ion with $\tan^{-1} (v_{\perp}/v_{\parallel}) \sim \pi /2$ is confined in the inward of the torus, but that losses in the outward of the torus. For another case of $\theta$ = 20 degree, the confinement of the co-going ion improves, and small amount of the counter-going ion is confined. Also, the almost trapped ion with $\tan^{-1} (v_{\perp}/v_{\parallel}) \sim \pi / 2$ is confined. However, for the last case of $\theta$ = 30 degree, the number of confined co-going ion decreases, and all counter-going ions loss. The number of the trapped ion increases due to the increased helical ripple, and the almost trapped ion is confined. Since the orbit property is sensitive to the finite-$\beta$ effect, the orbit following calculation for the finite-$\beta$ equilibrium will be important.

\section{Summary and outlook}

This paper studied the simple design of the stellarator with tilted TF coils. Coupling the tilted TF coil and axisymmetric PF coil, the stellarator field produces. With tracing the magnetic field line for the vacuum field, the formation of clear and nested flux surfaces are confirmed. The rotational transform is proportional to the tilting angle of the TF coil, $\theta$, and the magnetic shear is weak. The spectrum of the magnetic field on Boozer coordinate is studied. The stellarator with the tilted TF coils has a similar property to the quasi-isodynamic configuration, because the mirror ripple is a dominant component of the magnetic field spectrum. With changing the tilting angle of the TF coil, $\theta$, the configuration is scanned. For a case of a low tilting angle, $\theta$ = 10 degree, the rotational transform is small and the magnetic shear is very weak. In such a case, the drift width of the proton orbit is large, and then the co-going ions cannot be confined. A small amount of the counter-going ion in the inward of the torus can be confined. However, for another case of $\theta$ = 20 degree, the rotational transform increases, and the confinement of the co-going orbit improves. Also, the confinement of the trapped ion for the pitch angle, $\tan^{-1} (v_{\perp}/v_{\parallel}) \sim \pi /2$, improves. However, for the last case of $\theta$ = 30 degree, the confinement of the co-going ions degrades again.

The design of the simple stellarator in this paper is studied for only the vacuum field. Since the magnetic shear is weak, the magnetic field might be sensitive to the finite-$\beta$ effect. Studies including the finite-$\beta$ effect is an important subject. Also, in this paper, we discussed only the tilted TF coils according to a simple rule, that is, the TF coil tilts along the toroidal angle, $\phi$, direction. From a study of CNT, it was confirmed that the magnetic field is very sensitive to the small displacement of the coil set~\cite{Zhu2018}. Therefore, the optimization of the simple stellarator will be possible. Those studies are future subjects, and those will be reported in near future. 

\section*{Acknowledgments}
This work is supported by the NINS (National Institute of Natural Sciences) program of Strategic International Research Interaction Acceleration Initiative (Grant No. UFEX404), and State Key Laboratory of Advanced Electromagnetic Engineering and Technology (Grant No. AEET 2019KF002). Also, this work was partially supported by 'PLADyS', JSPS Core-to-Core Program, A. Advanced Research Networks.

\bibliography{soft2020}

\end{document}